\documentclass[11pt]{article}

\usepackage[english]{babel}
\usepackage{amsmath, latexsym}
\usepackage{amssymb}
\usepackage[authoryear, round]{natbib}

   \usepackage[dvips]{graphicx}
   \usepackage{rotating}

\newtheorem{theorem}{Theorem}[section]

\newtheorem{proposition}[theorem]{Proposition}

\newtheorem{corollary}[theorem]{Corollary}

\newlength{\captionwidth}
\begin{document}

\title{Incorporating exchange rate risk into PDs and asset correlations}

\author{%
Dirk Tasche\thanks{
E-mail:
dirk.tasche@gmx.net
}}

\date{December 2007}
\maketitle

\begin{abstract}
Intuitively, the default risk of a single borrower is higher when her or his assets and debt are denominated in different currencies. Additionally, the default dependence of borrowers with assets and debt in different currencies should be stronger than in the one-currency case. By combining well-known models by \citet{Merton1974}, \citet{GarmanKohlhagen}, and \citet{Vasicek2002} we develop simple representations of PDs and asset correlations that take into account exchange rate risk. From these results, consistency conditions can be derived that link the changes in PD and asset correlation and do not require knowledge of hard-to-estimate parameters like asset value volatility.
\end{abstract}


\section{Introduction}
\label{se:summary}

If borrowers have only assets that when liquidated generate
cash in a local currency different from the currency in which their
debt is due, their default risk will be higher than in the one currency
case, as a consequence of the additional exchange rate risk. The increase
in default risk is reflected both in higher probabilities of default (PDs) as well as in higher asset correlations
between the borrowers.

In this note, by modifying Merton's model of the default of a firm, we derive
some simple relations between the PDs without and with
exchange rate risk, between the borrowers' asset correlations without and with exchange
rate risk, and PDs and asset correlations when taking account of exchange rate risk.

In general, the formulae we derive include as parameters the borrowers' asset volatilities,
the exchange rate volatility, and the mean logarithmic ratio of the exchange rates at times 1 and 0.
However, assuming independence of the exchange rate and the borrowers' asset values as well as
zero mean logarithmic ratio of exchange rates at times 1 and 0 yields a relation between
the borrowers' asset correlation without and with exchange rate risk and the borrowers' PDs
without and with exchange rate risk that does not require knowledge of additional parameters
(see Equation \eqref{eq:relation}).
In the special case of borrowers with identical individual risk characteristics (= PDs), relation
\eqref{eq:relation} can be
stated as follows:
\begin{equation}\label{eq:simple}
    \frac{1-\varrho^\ast}{1-\varrho}\ = \ \frac{\Phi^{-1}(p^\ast)^2}{\Phi^{-1}(p)^2},
\end{equation}
where $p$ and $\varrho$ denote the original PD and asset correlation without exchange rate risk and
$p^\ast$ and $\varrho^\ast$ denote the PD and asset correlation when there is additional exchange
rate risk. Both \eqref{eq:simple} and \eqref{eq:relation} can be understood as consistency conditions
that should be satisfied when the risk parameters PD and asset correlation are to be adjusted
for incorporating exchange rate risk.

We describe in Section \ref{sec:just} the background of the model we use. In Section \ref{se:metho}, it is shown how the results are
derived from the model. The note concludes with a brief discussion of what has been reached.


\section{Background of the model}
\label{sec:just}

As in Merton's model for the default of a firm \citep{Merton1974}, we assume that $A(t)$, the borrower's asset value as a function of time, can be described by a geometric
Brownian Motion, i.e.
\begin{align}
d\,A(t) & \ = \ \mu\,A(t)\,d\,t + \sigma\,A(t)\, d\,W(t), \notag \\
\intertext{or, equivalently}
    A(t)& \ =\ A_0\,\exp\bigl((\mu-\sigma^2/2)\,t + \sigma\,W(t)\bigr), \quad t \ge 0, \label{eq:asset_value}
\end{align}
where $A_0$ is the asset value at time $0$ (today), $\mu$ is the drift of the asset value
process, $\sigma$ is its volatility, and $W(t), t \ge 0$ denotes a Standard Brownian Motion
that explains the randomness of the future asset values.

Similar to \eqref{eq:asset_value}, we assume that $F(t)$, the exchange rate of the two currencies
at time $t$, can be described as another geometric Brownian Motion \citep{GarmanKohlhagen}, i.e.
\begin{align}
d\,F(t) & \ = \ \nu\,F(t)\,d\,t + \tau\,F(t)\, d\,V(t), \notag \\
\intertext{or, equivalently}
    F(t) &\ =\ F_0\,\exp\bigl((\nu-\tau^2/2)\,t + \tau\,V(t)\bigr), \quad t \ge 0, \label{eq:FX_rate}
\end{align}
where $F_0$ is the exchange rate at time $0$, $\nu$ is the drift of the exchange rate
process, $\tau$ is its volatility, and $V(t), t \ge 0$ denotes another Standard Brownian Motion
that explains the randomness of the future exchange rates.

The Brownian Motions $V(t), W(t), t \ge 0$ are correlated with correlation parameter $r$,
i.e.
\begin{equation}\label{eq:corr_Brown}
    \mathrm{corr}[V(t)-V(s),\,W(t)-W(s)]\ = \ r, \quad 0 \le s < t.
\end{equation}
As in Merton's model of the default of a firm, the borrower defaults after one year (i.e.~$t=1$) if her or his asset value
by then has fallen below her or his level of due debt $D$. However, debt is due in a currency different from the
currency in which the asset value is denominated. Hence the asset value must be multiplied with the exchange rate
at time 1:
\begin{subequations}
\begin{equation}\label{eq:def_level}
    F(1)\,A(1)\ \le \ D.
\end{equation}
From an economic point of view, it is convenient to divide both sides of \eqref{eq:def_level} by $F_0$. This leads
to
\begin{equation}
	\label{eq:def_level_a}
F^\ast(1)\,A(1) \ \le \ D^\ast	
\end{equation}
with
\begin{equation}
	\label{eq:def_level_b}
\begin{split}	
F^\ast(t)&\ =\	\exp\bigl((\nu-\tau^2/2)\,t + \tau\,V(t)\bigr),\\
D^\ast&\ =\ D/F_0.
\end{split}
\end{equation}
\end{subequations}
The advantage of \eqref{eq:def_level_a} compared to \eqref{eq:def_level} is the fact that on the
one hand the debt
level is expressed as a value in the local currency of the borrower's assets with an exchange rate as observed today.
On the other hand, compared to the one currency case the volatility of the left-hand side of
\eqref{eq:def_level_a} is higher because it includes the factor $F^\ast(1)$ that reflects the
change of the exchange rate between today and time 1. This effect might be mitigated to some extent by the
difference of the interest rates in the two countries. For the purpose of this note, however, it is assumed that
mitigation by interest rates differences can be neglected. This assumption seems justified in particular when
the debt is composed of fixed rate loans or is short-term.

Taking the logarithm of both sides of \eqref{eq:def_level_a} and standardisation of
the random variable
\begin{equation}\label{eq:logassets}
    \log(A(1))\ =\ \log A_0 + \mu - \sigma^2/2 + \sigma\,W(1)
\end{equation}
lead to
\begin{equation}\label{eq:log}
    \frac{\nu - \tau^2/2 + \tau\,V(1)}{\sigma} + W(1) \ \le\ \frac{\log D - \log A_0 - \log F_0 - \mu + \sigma^2/2}{\sigma}.
\end{equation}
Define now $F = \nu - \tau^2/2 + \tau\,V(1)$, $A = W(1)$, and $c = \frac{\log D - \log A_0 - \log F_0
		- \mu + \sigma^2/2}{\sigma}$ to arrive at
\begin{subequations}
\begin{equation}\label{eq:diff_currency}
   \tfrac{F}{\sigma} + A \ \le \ c.
\end{equation}
In \eqref{eq:diff_currency},  $F \sim N(\nu, \tau^2)$ is the logarithmic ratio of the exchange rates at times 1 and
0 and is jointly normally distributed with $A \sim N(0,1)$. As a consequence from
\eqref{eq:corr_Brown}, the correlation of $F$ and $A$
is given by
\begin{equation}\label{eq:corr_FX}
    \mathrm{corr}[F,\, A]\ = \ r.
\end{equation}
\end{subequations}

Note that, due to the convexity of the exponential function, $\mathrm{E}[F^\ast(1)]= 1$ is \textbf{not} equivalent to
$\mathrm{E}[F] = 0$ but to $\mathrm{E}[F] = - \tau^2/2$. If $\mathrm{E}[F] = 0$ on the other hand, then $\mathrm{E}[F^\ast(1)]= \tau^2/2$.

\section{Model \& Results}
\label{se:metho}

\citet[][see also the references therein]{Vasicek2002} suggested to take model \eqref{eq:asset_value}
of a borrower's asset value as the basis for a model of the joint default behaviour of several
borrowers. As we consider the evolution in one period only, standardisation similar to the one
used for deriving \eqref{eq:diff_currency} shows that
the PDs $p_1$, $p_2$ of borrowers 1 and 2 can be characterised by the equation
\begin{equation}\label{eq:same_currency}
    A_i \ \le \ c_i \ = \Phi^{-1}(p_i),\ i=1,2
\end{equation}
where $(A_1, A_2) \sim N\left(\left(\begin{smallmatrix}0\\  0\end{smallmatrix}\right),
\left(\begin{smallmatrix}1 & \varrho\\ \varrho & 1\end{smallmatrix}\right)\right)$.
The numbers $c_1, c_2$ are called \emph{default thresholds}. The correlation $\varrho$ is called \emph{asset correlation}
because it can be interpreted as the correlation of the changes in asset values of the two borrowers. Equation \eqref{eq:same_currency}
does not yet include the case of assets and debt being denominated in different currencies.

Assume that borrowers 1 and 2 receive their revenues mainly from one country and, hence, have their assets denominated
in the local currency. Assume, however, that their debt is due in another currency. Combining the reasoning that led
to \eqref{eq:diff_currency} and \eqref{eq:corr_FX} with the reasoning for \eqref{eq:same_currency} then yields a description
of the joint default behaviour under exchange rate uncertainty as follows:
The default events for borrowers 1 and 2 are characterised  by the equation
\begin{subequations}
\begin{equation}\label{eq:diff_currency_2}
   \tfrac{F}{\sigma_i} + A_i \ \le \ c_i, \quad i=1,2
\end{equation}
where $(A_1, A_2)$ and $c_1,c_2$ are as in \eqref{eq:same_currency}\footnote{%
That $c_1, c_2$ should be the same as in \eqref{eq:same_currency} might be surprising at first glance. It follows, however,
from the fact that in \eqref{eq:def_level_a} the foreign currency debt amount has been converted to a local currency amount at
today's exchange rate. Additional uncertainty and higher default risk are then caused by the fact that the exchange rate in one year
may differ from today's rate.}.
$F$ is related to the annual change in the exchange rate between the both currencies (see
\eqref{eq:log}). Moreover, $F \sim N(\nu, \tau^2)$ is assumed
to be jointly normally distributed with $(A_1, A_2)$:
\begin{equation}\label{eq:corr_FX_2}
    \mathrm{corr}[F,\, A_i]\ = \ r_i, \quad i=1,2.
\end{equation}
$\sigma_i > 0$ denotes the volatility of the asset value process of borrower $i$ as in \eqref{eq:asset_value}.
\end{subequations}

Equation \eqref{eq:diff_currency_2} and the assumptions on the joint normal distribution of exchange rate and
asset values readily imply the following result
on the change in PD and asset correlation when exchange risk for the foreign currency denominated
risk is taken into account.

\begin{proposition}\label{pr:1}
\begin{subequations}
When the borrowers' assets and debt are denominated in different currencies, their probabilities
of default should be adjusted the following way:
\begin{equation}\label{eq:PD_FX}
    p_i^\ast \ = \ \mathrm{P}\bigl[\tfrac{F}{\sigma_i} + A_i \, \le \, c_i\bigr]\ = \
\Phi\left(\frac{c_i - \nu/\sigma_i}{\sqrt{\tau^2/\sigma_i^2 + 1 +2\,r_i\,\tau/\sigma_i}}\right).
\end{equation}
In \eqref{eq:PD_FX}, $c_i$ depends on the one-currency-only PD $p_i$ according to \eqref{eq:same_currency},
$\nu$ and $\tau$ denote the exchange rate drift and volatility according to \eqref{eq:FX_rate}, $\sigma_i$ stands
for the volatility of borrower $i$'s assets according to \eqref{eq:asset_value}, and $r_i$ is the correlation of
asset value changes and exchange rate changes according to \eqref{eq:corr_Brown}.

Additionally, the borrowers' asset correlation should be adjusted:
\begin{equation}\label{eq:corr}
    \varrho^\ast \ = \ \mathrm{corr}\bigl[\tfrac{F}{\sigma_1} + A_1,\, \tfrac{F}{\sigma_2} + A_2\bigr]
    \ = \ \frac{\varrho + r_1\,\tau/\sigma_1 + r_2\,\tau/\sigma_2 + \tau^2/(\sigma_1\,\sigma_2)}
            {\sqrt{\tau^2/\sigma_1^2 + 1 +2\,r_1\,\tau/\sigma_1}\,\sqrt{\tau^2/\sigma_2^2 + 1 +2\,r_2\,\tau/\sigma_2}},
\end{equation}
where $\varrho$ denotes the one-currency-only asset correlation of the borrowers according to \eqref{eq:same_currency}
and all other quantities are as in \eqref{eq:PD_FX}.
\end{subequations}
\end{proposition}
In order to apply adjustments to PD and asset correlation according to Proposition \ref{pr:1}, in particular estimates
of the asset volatilities $\sigma_1, \sigma_2$ must be provided. As in general the asset value processes cannot be directly observed,
such estimates are notoriously hard to obtain. Under additional simplifying assumptions, Proposition \ref{pr:1} nonetheless implies the following useful
consistency condition that might be taken into account when exchange rate risk adjustments are applied to PDs and asset correlations.

\begin{corollary}\label{co:1}
If the exchange rate $F$ and the asset values $(A_1, A_2)$ are uncorrelated (i.e.\ $r_1 = 0 = r_2$ in \eqref{eq:corr_FX_2}) and
the mean logarithmic ratio of the exchange rates at times 1 and
0 can be neglected (i.e.\ $\mathrm{E}[F] = \nu = 0$), then the following equation links up the adjusted PDs and asset correlations:
\begin{equation}\label{eq:relation}
    \varrho^\ast\,\frac{c_1}{\Phi^{-1}(p_1^\ast)} \,\frac{c_2}{\Phi^{-1}(p_2^\ast)}\ = \
    \varrho + \sqrt{\frac{c_1^2}{\Phi^{-1}(p_1^\ast)^2}-1}\,\sqrt{\frac{c_2^2}{\Phi^{-1}(p_2^\ast)^2}-1}.
\end{equation}
\end{corollary}

To see that Corollary \ref{co:1} follows  from Proposition \ref{pr:1}, solve \eqref{eq:PD_FX} for $\tau/\sigma_i$ and insert
the result into \eqref{eq:corr}.
Equation \eqref{eq:relation} can be generalised easily for $r_1, r_2 \not= 0$. Generalisation for the case $\nu \not= 0$, however, would again involve asset volatilities and exchange rate volatility as additional parameters.

Considerable simplification can be reached in \eqref{eq:relation} when it is assumed that the individual risk characteristics of the
borrowers are identical (homogeneous portfolio assumption), i.e.\
\begin{equation}\label{eq:assumption}
    p_1 = p_2 \quad \text{and} \quad p_1^\ast = p_2^\ast.
\end{equation}
In this case \eqref{eq:relation} can be equivalently written as \eqref{eq:simple}. Figure \ref{fig:1} presents
for some fixed combinations of original PDs and original asset correlations the relationship of adjusted PD and adjusted asset correlation. Note that \eqref{eq:PD_FX} implies that $p_i^\ast > p_i$ as long as $p_i < 0.5$ if $\nu = 0$ and $r_i = 0$. Hence the curves in Figure \ref{fig:1} start in the original PDs. Moreover, all the curves converge towards correlation 1 when the original PD approaches 0.5. In general, the curves are strictly concave such that the asset correlations grow over-proportionally as long as the adjusted PDs are small, and under-proportionally when the adjusted PDs are high (close to 0.5).

\setcounter{figure}{0}
\refstepcounter{figure}
\begin{figure}[ht]
\centering
  \parbox{14.0cm}{Figure \thefigure:
  \emph{Adjusted asset correlation as function of adjusted PD according to \eqref{eq:simple}. Original PD and original asset correlation are fixed.}}
\label{fig:1}\\[2ex]
\begin{turn}{270}
\resizebox{\height}{14.0cm}{\includegraphics[width=14.0cm]{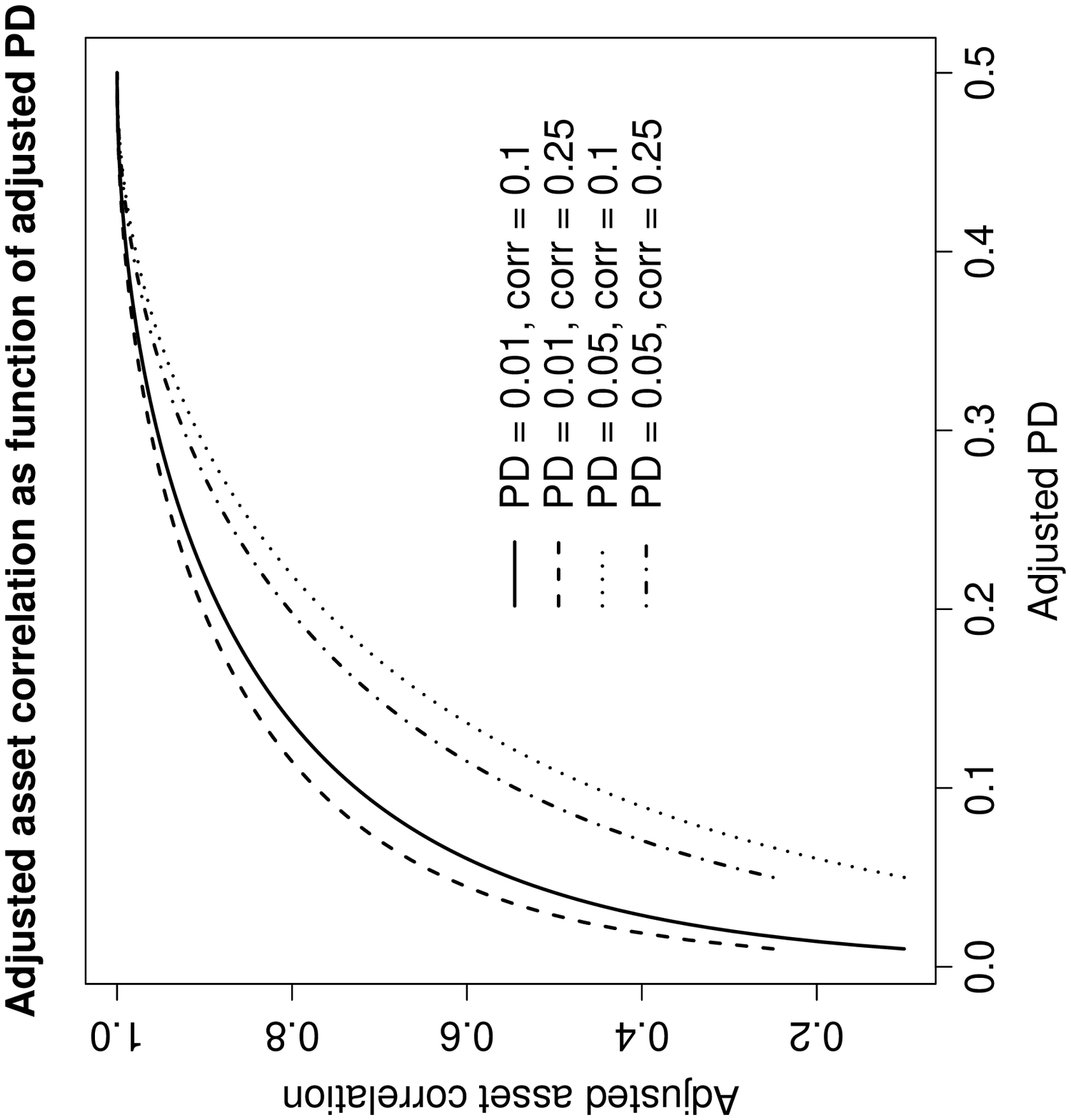}}
\end{turn}
\end{figure}

\section{Conclusions}
\label{se:conc}

Taking recourse to well-known models by \citet{Merton1974}, \citet{GarmanKohlhagen}, and \citet{Vasicek2002}
we have derived a simple model for incorporating exchange rate risk into the PDs and asset correlations of borrowers whose assets and debt are denominated in different currencies. On principle, the results can be used to derive values of PDs and asset correlations that are adjusted to take account of exchange rate risk. However, this requires knowledge of parameters like the borrowers' asset value volatility that are hard to estimate. Another type of result is potentially more useful because -- thanks to additional assumptions like the independence of exchange rate and asset value processes -- it represents a consistency condition for the joint change of PDs and asset correlations when exchange rate risk is taken into account. This latter result does not involve asset value volatilities or other inaccessible parameters.



\begin{thebibliography}{3}
\providecommand{\natexlab}[1]{#1}
\providecommand{\url}[1]{\texttt{#1}}
\expandafter\ifx\csname urlstyle\endcsname\relax
  \providecommand{\doi}[1]{doi: #1}\else
  \providecommand{\doi}{doi: \begingroup \urlstyle{rm}\Url}\fi

\bibitem[Garman and Kohlhagen(1983)]{GarmanKohlhagen}
M.~B.\ Garman and S.~W. Kohlhagen.
\newblock Foreign currency option values.
\newblock \emph{Journal of International Money and Finance}, 2:\penalty0
  231--237, 1983.

\bibitem[Merton(1974)]{Merton1974}
R.~C. Merton.
\newblock On the {P}ricing of {C}orporate {D}ebt: {T}he {R}isk {S}tructure of
  {I}nterest {R}ates.
\newblock \emph{Journal of Finance}, 29\penalty0 (2):\penalty0 449--470, 1974.

\bibitem[Vasicek(2002)]{Vasicek2002}
O.~A. Vasicek.
\newblock The distribution of loan portfolio value.
\newblock \emph{{RISK}}, 15:\penalty0 160--162, 2002.

\end{thebibliography}


\end{document}